\documentclass[
]{ceurart}

\usepackage{graphicx}
\usepackage{subcaption}
\usepackage[justification=centering]{caption}

\usepackage{amsmath,amssymb,amsfonts}
\usepackage[linesnumbered,algoruled,boxed,lined]{algorithm2e}
\usepackage{algorithmic}
\usepackage{textcomp}
\usepackage{hyperref}
\usepackage{listings}
\usepackage{multirow}
\usepackage{pifont}
\usepackage{courier}
\usepackage{blindtext}
\usepackage{enumitem}
\usepackage[switch]{lineno} 
\usepackage{booktabs}
\usepackage{mathtools}
\usepackage{xparse}% http://ctan.org/pkg/xparse
\usepackage{longtable}
\usepackage{svg}
\usepackage{makecell}
\usepackage{arydshln}
\usepackage{wrapfig} % for wrapfig
\usepackage[normalem]{ulem}
\usepackage{dirtytalk}

\usepackage{amsmath}
\usepackage{mathrsfs}
\usepackage{phoenician}
\usepackage{xspace}

\newcommand{\ra}[1]{\renewcommand{\arraystretch}{#1}}
\newcommand{\first}[1]{#1}
\newcommand{\second}[1]{#1}
\newcommand{\del}[1]{\textcolor{BrickRed}{}}

\newcommand{\pp}{\reflectbox{$\mathrm{P}$}\mkern-7mu\mathrm{P}}

\newcommand{\ppm}{$\pp$\xspace}

\definecolor{codegreen}{rgb}{0,0.6,0}
\definecolor{codegray}{rgb}{0.5,0.5,0.5}
\definecolor{codepurple}{rgb}{0.58,0,0.82}
\definecolor{backcolour}{rgb}{0.95,0.95,0.92}

\lstdefinestyle{mystyle}{
    backgroundcolor=\color{backcolour},   
    commentstyle=\color{codegreen},
    keywordstyle=\color{magenta},
    numberstyle=\tiny\color{codegray},
    stringstyle=\color{codepurple},
    basicstyle=\ttfamily\footnotesize,
    breakatwhitespace=false,         
    breaklines=true,                 
    captionpos=b,                    
    keepspaces=true,                 
    numbers=left,                    
    numbersep=5pt,                  
    showspaces=false,                
    showstringspaces=false,
    showtabs=false,                  
    tabsize=2
}

\begin{document}

\conference{BigHPC2024: Special Track on Big Data and High-Performance Computing, co-located with the 3\textsuperscript{rd} Italian Conference on Big Data and Data Science, ITADATA2024, September 17 -- 19, 2024, Pisa, Italy.}

\copyrightyear{2024}
\copyrightclause{Copyright for this paper by its authors.
  Use permitted under Creative Commons License Attribution 4.0
  International (CC BY 4.0).}

\title{miniLB: A Performance Portability Study of Lattice-Boltzmann Simulations}

% \tnotetext[1]{You can use this document as the template for preparing your
%   publication. We recommend using the latest version of the ceurart style.}

\author[1]{Luigi Crisci}[%
% orcid=0000-0002-0877-7063,
email=lcrisci@unisa.it,
% url=https://yamadharma.github.io/,
]
\cormark[1]
\address[1]{University of Salerno,
  Via Giovanni Paolo II 132, 80084, Fisciano, Italy}

\author[1]{Biagio Cosenza}[%
% orcid=0000-0001-7116-9338,
email=bcosenza@unisa.it,
% url=https://kmitd.github.io/ilaria/,
]

\author[2]{Giorgio Amati}[]

\author[2]{Matteo Turisini}

% \author[3]{Filippo Spiga}[]
% \author[4]{Georgios Markomanolis}[]

\address[2]{CINECA,
 Via dei Tizii, 6b, 00185 Roma, Italy}
%\address[3]{NVIDIA,
%  Cambridge, England}
% \address[4]{AMD,
%   Grenoble, France}
  
\cortext[1]{Corresponding author.}
% \fntext[1]{These authors contributed equally.}

\begin{abstract}
The Lattice Boltzmann Method (LBM) is a computational technique of Computational Fluid Dynamics (CFD) that has gained popularity due to its high parallelism and ability to handle complex geometries with minimal effort. Although LBM frameworks are increasingly important in various industries and research fields, their complexity makes them difficult to modify and can lead to suboptimal performance.
This paper presents \emph{miniLB}, the first, to the best of our knowledge, SYCL-based LBM mini-app. 
\emph{miniLB} addresses the need for a performance-portable LBM proxy app capable of abstracting complex fluid dynamics simulations across heterogeneous computing systems. 
We analyze SYCL semantics for performance portability and evaluate \emph{miniLB} on multiple GPU architectures using various SYCL implementations. 
Our results, compared against a manually-tuned FORTRAN version, demonstrate effectiveness of \emph{miniLB} in assessing LBM performance across diverse hardware, offering valuable insights for optimizing large-scale LBM frameworks in modern computing environments.
\end{abstract}


\maketitle

\section{Introduction}
In High-Performance Computing (HPC), \textit{mini-apps} (or \textit{proxy-apps}) are simplified codes that allow application developers to share and analyze important key features of large applications without forcing users to assimilate large and complex code bases. 
Mini-apps are often used as abstract models to evaluate performance and assess performance, portability, and performance portability (\ppm). 
Mini-app can also capture programming methods and styles that drive requirements for algorithms, compilers, and other toolchain elements.  Developing mini-apps for relevant use cases is an important challenge in pushing the boundaries of HPC application performance. Important projects such as the Exascale Proxy Applications Project \cite{proxy} aim to improve the quality of proxies produced by the Exascale Computing Project by defining standards for documentation, building and testing systems, performance models and evaluations, and templates and best practices for proxy developers to help meet these standards.

In recent years, the Lattice Boltzmann Method (LBM) has further strengthened its position as a valuable tool in the field of computational fluid dynamics \cite{lb_industry}. LBM has attracted increasing interest in many industries and research organizations due to its high parallelism efficiency and ability to discretize complex geometries with little effort. This has led to the development of large frameworks, typically focused on specific LBM domains, with extremely complex and large code bases \cite{openlb,walberla,palabos}. However, LBM frameworks have not evolved with the evolution of (massively parallel and distributed) computing systems, resulting in complex codebases that are very difficult to modify. 
Unfortunately, no mini-app for LBM can efficiently abstract the problem while providing hints for performance tuning and optimization.

This paper proposes the first mini-application for LBM, with an implementation in SYCL that allows not only performance evaluation but also performance portability on modern heterogeneous computing systems.
Specifically, this paper makes the following contributions:
\begin{itemize}
    \item The, to the best of our knowledge, first performance portable, tunable, SYCL-based Lattice-Boltzmann mini-app, translated from an original FORTRAN implementation, capable of targeting a wide range of multicore CPUs, GPUs, and accelerators.
    \item A performance portability analysis of SYCL semantics, including Unified Shared Memory, range, and ND-range kernels using a performance portability metric.
    \item An experimental evaluation of \emph{miniLB} on NVIDIA V100S, AMD MI100, and  Intel Max 1100 GPUs, using multiple SYCL implementations and different SYCL semantic combinations, compared to the manually-tuned FORTRAN version using different parallelism paradigms and compilers.
\end{itemize}

\section{Background and Related Work}
\paragraph{Lattice Boltzmann Method}
The Lattice Boltzmann Method (LBM) emerged in the late 1980s as an evolution of lattice gas cellular automata. It has since found numerous applications across various complex flow problems, from fully developed turbulence to micro and nanofluidics, and even quark-gluon plasmas. 

The core concept of LBM is to solve a simplified Boltzmann kinetic equation for a set of discrete distribution functions, known as populations, \( f_i(\mathbf{x}; t) \). These functions represent the probability of finding a particle at position \( \mathbf{x} \) and time \( t \), with a discrete velocity \( \mathbf{v} = \mathbf{c}_i \). The discrete velocities are selected to ensure sufficient symmetry, thereby satisfying the mass, momentum, and energy conservation laws of macroscopic hydrodynamics and maintaining rotational symmetry. 
Figure \ref{eq1} illustrates the lattices used for 2D LB simulations, featuring a set of nine discrete velocities (D2Q9). Instead of directly solving the Navier-Stokes equations ($\rho\vec{u}$), LBM solves a kinetic equation storing nine populations for each grid point, corresponding to different \( \mathbf{c} \) velocity directions, including \( \mathbf{c} = (0, 0) \). It is not necessary to store derived quantities like velocity and density.

In its compact form, the main LB equation is as follows:
\begin{equation}\label{eq1}
   f(\vec{x}+\vec{c_i}, t + 1) - f_i(\vec{x}; t) = 
   -\omega(f_i(\vec{x}; t) - f_i^{eq}(\vec{x}; t)) + S_i, \;\;\;i \in [0,b+1] 
\end{equation}
where $\vec{x}$ and $\vec{c_i}$ are position and velocity vectors in ordinary space, $f_i^{eq}$ is the equilibrium distribution function, $S_i$ is a source term for the fluid interaction with external (or internal) sources.
The local equilibria are provided by a lattice truncation, to the second order in the Mach number $M = u / c_s $, of the Maxwell-Boltzmann distribution, where $c_s$ is the lattice sound speed.

\begin{equation}
    f_i^{eq}(\vec{x}, t) = w_i \rho (1 + u_i + q_i)
\end{equation}
where $w_i$ is a set of weights normalized to the unit and,
\begin{equation}
    u_i = 3 \frac{\vec{u} * \vec{c_i}}{c_s} \;\;\;\;\;\;\;\;\;\;\;\;\;\;\; q_i = 9 \; 
    \frac{(\vec{u} * \vec{c_i})^2}{2c_s^2} - 3\; \frac{u^2}{2c^2_s}
\end{equation}
where the left term is linear in velocity, the right one quadratic.
The equation \ref{eq1} represents two key processes: the \textit{collision step} (right-hand side), where the populations locally relax towards equilibrium, and the streaming step (left-hand side), where the populations are propagated to neighboring locations at \(\mathbf{x} + \mathbf{c}_i\) at time \(t + 1\). This scheme can be demonstrated to reproduce the Navier-Stokes equations for an isothermal, quasi-incompressible fluid in terms of density and velocity \cite{kruger}.

% Some LB frameworks are LIST AND CITATION....
% However, those are not micro framework BLABLA
% Other works explores SYCL on LB CITAZIONE, but it's not a mini-app.
In literature, several large Lattice Boltzmann frameworks exist, e.g. OpenLB\cite{openlb}, waLBerla\cite{walberla}, Palabos\cite{palabos}. However, such frameworks usually present very large and complex code bases, which makes it difficult to experiment with for research purpose. Our work aims at staying as simple as possible while providing a playground for experimenting with SYCL-specific or LB-specific optimizations while providing good performance on multiple hardware. 
Other attempts to verify the performance of heterogeneous programming models on LB methods have been proposed in the literature: Ding et.al. \cite{app-lb-sycl-kokkos} explore the performance of the SYCL and Kokkos programming model on an LB application, showing performance pitfalls of both implementations. However, their work does not focus on evaluating single SYCL features like miniLB but rather the raw application performance. Moreover, their analysis only focuses on NVIDIA GPUs, while we examine performance portability across multiple vendors.

\paragraph{SYCL}
SYCL \cite{sycl-specification} is a royalty-free, cross-platform C++ abstraction layer that enables developers to write code for multiple heterogeneous devices, such as CPUs, GPUs, and FPGAs, in a convenient and performance-portable way. SYCL enhances the C++ programming language by adding abstractions for managing heterogeneous computing within ISO C++, aiming to align closely with the core language specifications. Originally designed to map onto OpenCL, the third revision of the SYCL 2020 specification allowed for custom backends, like NVIDIA CUDA, AMD HIP, OpenMP, and others.  Key implementations of SYCL include Intel's OneAPI Data-Parallel C++ \cite{dpcpp} and AdaptiveCPP \cite{adaptivecpp}, along with several other smaller-size implementations \cite{triSYCL,neoSYCL, simsycl}. 
The versatility of SYCL has led to various extensions for specific heterogeneous computing scenarios, including distributed computing \cite{celerity}, real-time energy optimization \cite{syngergy}, and approximate computing \cite{syprox}. 

\paragraph{Performance Portability}
% HPC systems are continually advancing towards innovative architectural designs. With the expanding variety of hardware in modern HPC systems, it is imperative to develop application code that can operate efficiently across different devices.
% % In such a context, evaluating an application's efficiency solely based on specific hardware is insufficient. It is vital to establish a quantitative metric that encompasses both the performance achieved and the range of devices on which the application can execute.
% Consequently, the concept of performance portability has gained prominence in scholarly discussions, capturing two critical dimensions: the achievement of a benchmark level of performance on designated platforms and the capability to execute an application on various hardware architectures.
As HPC systems evolve with diverse hardware architectures, developing efficient, cross-device application code becomes crucial. This has led to the rise of "performance portability" in academic circles, which measures both an application's ability to meet performance benchmarks on specific platforms and its capacity to run across various hardware configurations.

% The Department of Energy (DoE) and the European Technology Platform for High-Performance Computing (ETP4HPC) have underscored the significance of performance portability in HPC.
However, a universally accepted definition is absent. 
In our research, we embrace the definition of performance portability by Pennycook et al. \cite{pennycook2016metric}: \textit{"A measurement of an application's performance efficiency for a given task that can be successfully executed on all platforms within a specified set." }
The formula to quantify performance portability is presented in Equation \ref{eq:PP_metric}:

\begin{equation} \label{eq:PP_metric}
\pp(a,p,H)=
\begin{cases} 
    \frac{|H|}{\sum_{i\in H}\frac{1}{e_i(a,p)}}, & \text{if platform $i$ is supported for all $i \in H$} \\
    0, & \text{otherwise}
\end{cases}
\end{equation}

Here, $a$ represents the application, $p$ denotes the problem addressed by $a$, and $H$ signifies the set of target hardware. The performance portability metric \ppm is defined as the harmonic mean of the application's performance efficiency $e_i(a,p)$ over the set of hardware $H$.

Pennycook et al. \cite{pennycook2016metric} highlight various methods for calculating application performance efficiency, specifically: \textit{architectural efficiency}, which measures achieved performance as a fraction of peak hardware performance; and \textit{application efficiency}, which measures achieved performance as a fraction of the best-observed performance against the most optimized native implementation.

\section{miniLB Overview}

\subsection{Computational Description}

\emph{miniLB} is a bidimensional computational fluid dynamic code for single-phase incompressible flows, with nine discrete velocities (D2Q9 using CFD jargon).
It is a downsizing of a 3D \texttt{FORTRAN90} \texttt{MPI+OpenACC} full application, developed by CINECA \cite{Falcucci2021, AMATI2021101447}.
\emph{miniLB} is written in C++20 and SYCL, a single-source abstraction layer for heterogeneous computing. \emph{miniLB} has no external dependencies and uses no SYCL compiler-specific feature, allowing it to run out-of-the-box on every platform with any SYCL compiler.
The code is open-source and available on GitHub\footnote{https://github.com/Luigi-Crisci/miniLB}. \\

\noindent \emph{miniLB} implements a \textit{fused} approach \cite{fused1,fused2}, where the \textit{collision} and \textit{streaming} operation are performed in a single kernel. In this approach, the app holds a pre-collision population $f^{pre}$ and a post-collision population $f^{post}$: at time $t$, input values read through a scattered read from $f^{pre}$, and post-collision results are written in $f^{post}$. Finally, the two populations are swapped at time $t+1$. 
Populations are stored using a Structure-of-Array (SoA) layout, with a unit-strided vector for each population $f_i$.
\noindent \emph{miniLB} has been designed to be highly and easily tunable to measure SYCL performance in a wide range of scenarios, with a total of 96 possible configurations, highlighted below in section \ref{4_sycl_porting}. 
\subsubsection{Numerical precision}
\emph{miniLB} supports four different numerical precisions to control how data is computed and stored:(i) \texttt{Single}: quantities are stored in single precision and all floating point operations are performed in single precision. (ii) \texttt{Double}: quantities are stored in double precision and all floating point operations are performed in double precision. (iii) \texttt{Mixed1}: quantities are stored in half precision and all floating point operations are performed in single precision.  (iv) \texttt{Mixed2}: quantities are stored in single precision and all floating point operations are performed in double precision.

The \texttt{Single} precision option is the default one. A more comprehensive analysis of LBM numerical precision has been explored in \cite{lbm_precision}.

\subsection{FORTRAN-based Parallelization} \label{3_fortran_par}
The original FORTRAN app uses a \textit{directive-based} parallelization that minimize code refactoring from CPUs to GPUs implementation. It implements several programming models: \texttt{OpenACC} pragmas, \texttt{OpenMP Offload}, and the FORTRAN built-in operator \texttt{DOCONCURRENT}. 

While GPU vendors provide native compilation toolchains for some of those programming models, not all of them are supported by each vendor: for example, \texttt{OpenACC} supports only NVIDIA GPUs and AMD through the CRAY compiler, while \texttt{DOCONCURRENT} is not supported on AMD discrete GPUs.
Furthermore, to achieve optimal performance on specific hardware, users must compile their programs using the proprietary vendor compiler (e.g. nvFORTRAN for NVIDIA, amdclang for AMD, ifx for Intel). This requirement increases fragmentation and adds complexity, as it necessitates testing with additional toolchains.

\subsection{Use Cases}

\begin{figure}
    \centering
    \begin{subfigure}{0.49\linewidth}
        \centering
        \includegraphics[width=0.47\linewidth]{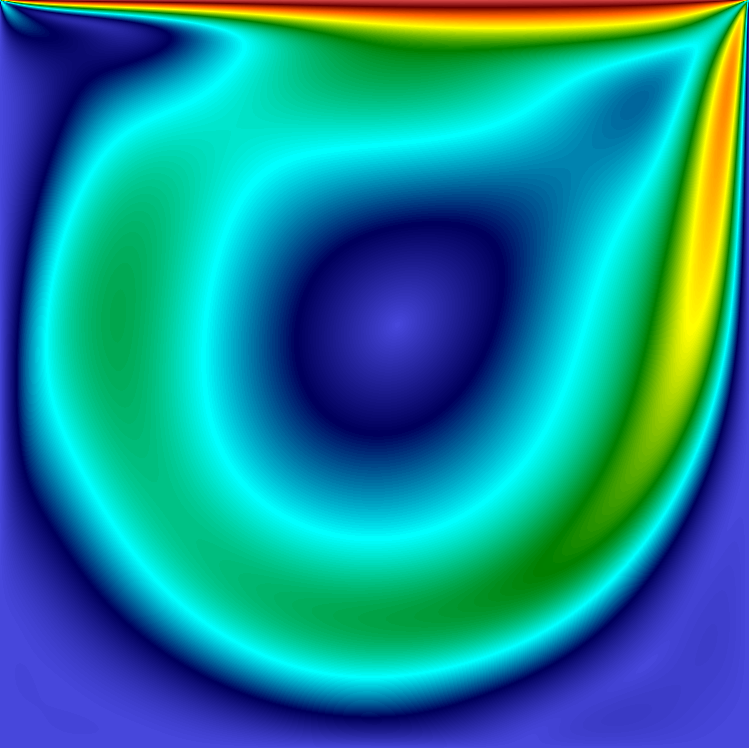}
        \caption{Lid-Driven cavity}
        \label{ldc}
    \end{subfigure}\hfill
    \begin{subfigure}{0.49\linewidth}
        \centering
        \includegraphics[width=\linewidth]{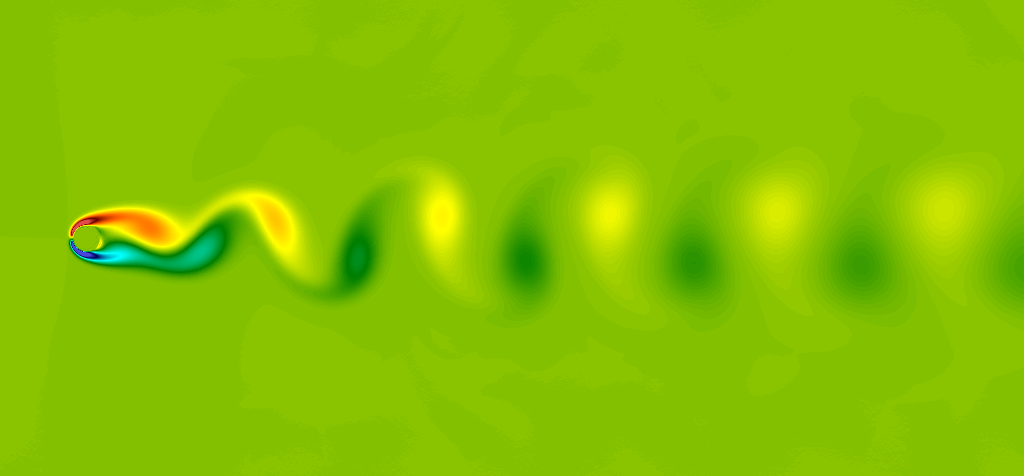}
        \caption{Von Karman Street}
        \label{vks}
    \end{subfigure}
    \caption{Use case example outputs using the .vtk file with ParaView}
    \label{fig:use-case}
\end{figure}

\emph{miniLB} supports three classic CFD benchmarks: Lid-Driven Cavity (LDC), Taylor-Green Vortex, and Von Karman Street (VKS). 
The app also produces VTK output files for offline visualization with tools like ParaView \cite{ParaView}. In this paper, we focus on the latter and the former: LDC and VKS.

\paragraph{Lid-Driven Cavity}
The LDC problem involves a square or rectangular cavity closed on all sides. The top lid of the cavity moves at a constant velocity, while the other three walls remain stationary. This setup generates a complex flow pattern within the cavity, characterized by the following: \textit{(a) } no-slip boundary conditions: all walls, including the moving lid, have a no-slip boundary condition, meaning the fluid velocity relative to the wall is zero;
\textit{(b)} driven flow: the movement of the top lid at a constant tangential velocity $u_0$ drives the flow within the cavity.
Figure \ref{ldc} shows an example output.

\paragraph{Von Karman Street} The VKS occurs when fluid flows past a cylindrical object and the flow separates alternately from either side of the object, creating a pattern of vortices in the wake. This phenomenon is characterized by: \textit{(a)} Periodic Vortex Shedding: alternating vortices are shed from opposite sides of the cylinder, creating a staggered array of vortices downstream. \textit{(b)} Flow Regimes: the flow pattern depends on the Reynolds number, which is a dimensionless number representing the ratio of inertial forces to viscous forces in the flow.
A visual example is given in figure \ref{vks}.

\section{SYCL Porting} \label{4_sycl_porting} % switch a Porting per enfatizzare il fatto che e' un porting
In this section, we analyze the principal SYCL features used during the porting and the available \emph{miniLB} configurations.

\subsection{Kernel Parallelism} 
%xxx mapping SYCL, parallel for, nd range, range
SYCL uses \lstinline{parallel_for} to declare parallel code regions. In particular, SYCL offers two variants of it: \texttt{range} and \texttt{NDrange}. The \texttt{range} is the simplest as it requires the user to specify only the iteration space. This allows the runtime to select the most appropriate number of threads depending on the target device without user intervention. On the other hand, \texttt{NDrange} allows to manually tweak the local iteration space (e.g. local workgroup size on OpenCL), allowing more fine-grained optimization but also requiring the user to manually optimize the local size to match the current device.
\emph{miniLB} kernels support both \texttt{range} and \texttt{NDrange}. 
The app defaults to \texttt{range} parallel for, but the latter can be activated by setting the \lstinline{-DBGK_SYCL_ND_RANGE} compile-time parameter. 
Currently, the app only supports the tweaking of the collide and stream kernel size through the \lstinline{-DBGK_SYCL_ND_RANGE_[X,Y]_SIZE} parameter at compile-time. 
 
\subsection{Data Management and Access}
% SYCL programming model provides two ways of handling data: the \textit{Buffer-Accessor (BA)} and the \textit{Unified Shared Memory} (USM) memory management interfaces. 
% With the former, the user wraps data into multi-dimensional \lstinline{buffer} objects and declares the data access pattern within a kernel with an \lstinline{accessor} object. 
% On the other hand, USM is a lower-level pointer API in which memory allocations and deallocations follow the malloc/free C paradigm. 

\emph{miniLB} uses Unified Shared Memory for data management instead of the \textit{Buffer-Accessor (BA)} paradigm as it shows more reliable and stable performance \cite{sycl-bench}.
USM provides three allocation kinds: \lstinline{malloc_device} are allocated directly on the target device, \lstinline{malloc_host} allocates host page-locked memory accessible from both host and device, and \lstinline{malloc_shared}, which are shared between devices using an automatic memory migration system. \\
\emph{miniLB} implements two memory management backends, one based on \lstinline{malloc_device} and \lstinline{malloc_host} and one using \lstinline{malloc_shared}, controllable via the compile-time parameter \lstinline{-DBGK_SYCL_MALLOC_SHARED}. 
In the former, \emph{miniLB} stores a device and host pointer for each population and manually migrates memory from the host and device. The host device is pinned because this increases the bandwidth with some GPU architectures \cite{pinned-memory}; in the latter, a single pointer is stored and the SYCL runtime handles the memory migration.
% When using shared allocations, SYCL allows the user to hint the runtime that a certain portion of your data is going to be accessed on the target device, thus allowing the runtime to migrate the data before the kernel execution. miniLB supports prefetching through the \lstinline{-DBGK_SYCL_ENALBE_PREFETCH} compile-time parameter \\
In addition, the parameter \lstinline{-DBGK_SYCL_ENALBE_PREFETCH} enables hints to prefetch memory on the host/device to the SYCL runtime. \\

Multi-dimensional data are defined with \emph{MDspan} \cite{mdpsan}. It is a lightweight, non-owning view that allows a piece of memory to be treated as a multi-dimensional entity. 
\emph{MDspan} allows us to define the extents (i.e. the number of dimensions and sizes), the layout (e.g. row-major, column-major, etc.), and the data accessor (i.e. how to translate the pointer/index pair to a memory location).  
\emph{miniLB} stores a two-dimensional \emph{MDspan} for both host and device to reduce the view construction overhead. In addition, a compile-time parameter \lstinline{-DBGK_SYCL_LAYOUT_[RIGHT|LEFT]} switches the data layout to row-major or column-major respectively. \emph{miniLB} defaults to column-major as it is the default layout in the original FORTRAN application.

\subsection{Task Scheduling}
To submit tasks, a SYCL user needs to create a \lstinline{queue} that binds to a specific device. SYCL supports both out-of-order and in-order submissions. With the former, submitted tasks are executed without a defined order, allowing for parallel kernels execution.
% If the kernel uses the \textit{BA} memory API, dependencies between kernels are automatically tracked. On the other hand, with USM the user needs to explicitly define dependencies by using the \lstinline{depends_on} function.
With USM, data dependencies between kernels must be explicitly tracked.
With in-order submission, instead, kernels are executed in FIFO order. This hinders the ability to parallelize kernel executions, but it removes the overhead of dependency tracking. \\
\emph{miniLB} supports both queue configuration, controlled by the parameter \lstinline{-DBGK_SYCL_IN_ORDER_QUEUE}.

% tabella per riassumere le combinazioni di configurazioni

% Tabella con AI e FP da ligen
% Tabella PP su device peak performance

\begin{table*}
    \ra{1.3}
      \centering
      \resizebox{0.8\columnwidth}{!}{%
      \begin{tabular}{@{}ccc@{}} \toprule
Feature                & Description & Options \\ \midrule
Precision & Kernel's floating point precision & \texttt{Single, Double, Mixed1, Mixed2} \\
Queue & SYCL queue & \texttt{Out-of-order, In-order} \\
USM & SYCL USM data management backend & \texttt{Device, Shared, Shared + Prefetch} \\
Layout & Data representation layout & \texttt{Row-major, Column-major} \\
Ranges & SYCL kernel types & \texttt{Range, NDRange} \\
\bottomrule
\end{tabular}%
}
\caption{\emph{miniLB} tuning configurations}
\label{tab:tuning_conf}

\end{table*}

\section{Experimental evaluation}

\subsection{Experimental Setup}
We evaluated \emph{miniLB} on three GPUs from three principal
GPU vendors, i.e. NVIDIA Tesla V100S, AMD MI100, and Intel Max 1100. 
We tested the app on two use cases: Lid-Driven Cavity (LDC) and Von Karman Street (VKS), using a $4096 * 4096$ grid with Reynolds number $Re = 10000$ and with 100.000 timesteps. 
As performance metric, we use MLUPs (Mega Lattice Update per second) that indicates how many millions of gridpoints are updated each second.

For the SYCL implementations, we chose AdaptiveCpp (commit sha a3c5c9d), and DPC++(commit sha ea0c067), where both support all three architectures.
\first{For AdaptiveCpp, we target the \textit{generic} backend, which can target every hardware through an integrated JIT compiler.}
For the FORTRAN version, we used NVHPC 24.5 for NVIDIA, amdclang ROCM 6.0 for AMD, and IFX 2024.0.0 for Intel.
To reduce the number of combinations in the tuning space,  this paper explores all combinations except \texttt{Mixed 1} and \texttt{Mixed 2} precision, SYCL out-of-order queues, row-major layout, and the \texttt{Shared+Prefetch} configuration.
% The \texttt{Shared + Prefetch} is not considered in this evaluation as it does not show the expected performance due to synchronization issues. %TOOD Troppo onesto??

\subsection{Use Case Evaluation}

\begin{figure}
    \centering
    \includegraphics[width=\linewidth]{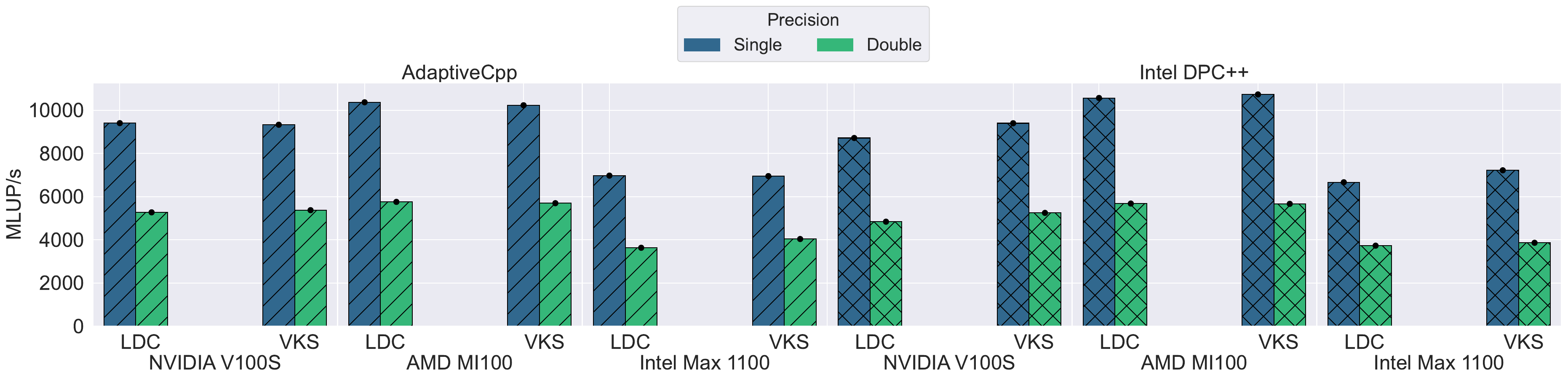}
    \caption{Best SYCL configuration Million Lattice Update per seconds (MLUPs) for Lid-Driven Cavity and Von Karmann Street use cases}
    \label{fig:mlups}
\end{figure}

\begin{figure}
    \centering
    \includegraphics[width=\linewidth]{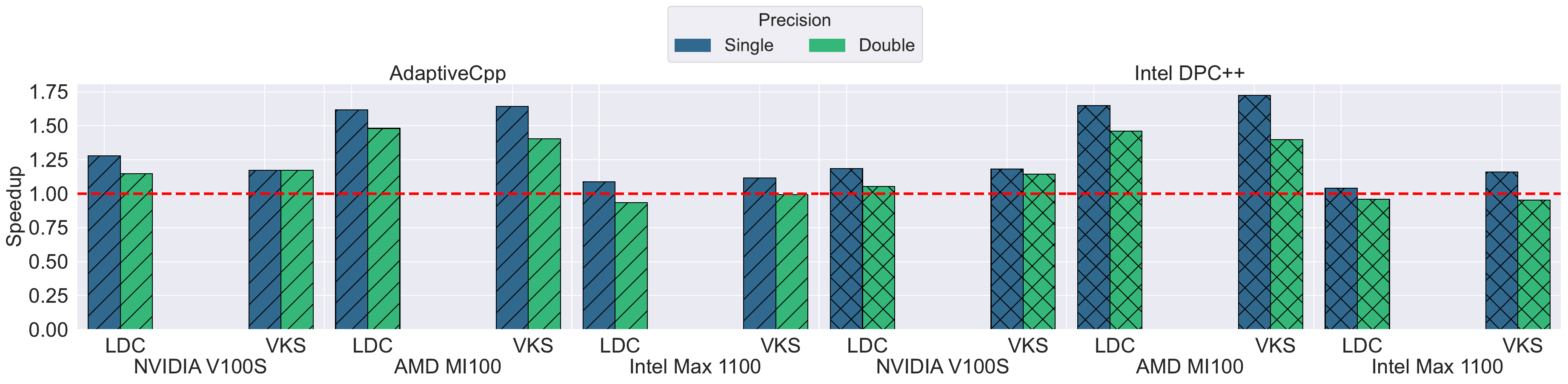}
    \caption{Best SYCL configuration speedup normalized over the Fortran best implementation}
    \label{fig:speedup-fortran-best}
\end{figure}

Figure \ref{fig:mlups} shows the achieved SYCL performance, while figure \ref{fig:speedup-fortran-best} shows the SYCL performance normalized to the best FORTRAN implementation on a given hardware \first{(e.g. OpenACC on LDC-NVIDIA, OpenMP on AMD-VKS)}. 
For each use case and hardware combination, the result of the best SYCL configuration is shown (e.g. for DPC++ on the NVIDIA V100S with the LDC use case, we pick \texttt{NDrange} + \texttt{device} allocation).
Both AdaptiveCpp and DPC++ show similar performance on every hardware on both use cases.
The SYCL version outdo FORTRAN on almost every platform and precision with both compilers, achieving an average of 1.16x speedup on the NVIDIA V100S, 1.03x on Intel Max 1100, and 1.54x on AMD MI100. 
An interesting case is AMD: while the results are similar to the ones obtained on other platforms, the speedup over FORTRAN is way higher(e.g. 1.55x compared to 1.16x on NVIDIA with AdaptiveCpp). 
To find the cause of this speedup, we profiled both the FORTRAN and SYCL application on single precision using OmniPerf \cite{omniperf}. We found that \texttt{amdclang} performs an aggressive thread coarsening, reducing the iteration space by a factor of 30.
This results in a significantly higher L1I and L1D cache misses by duplication, 26 miss per wave against the 0.04 miss per wave of the SYCL implementation. Moreover, the FORTRAN version achieves 53\% more L2 cache misses, severely limiting application bandwidth. Similar considerations apply to the double precision too. 

% Figure \ref{fig:mlups} shows the achieved SYCL performance. For each use case and hardware, the result of the best SYCL configuration is shown (e.g. for DPC++-NVIDIA-LDC, we pick NDrange + Device allocation).

% More L1D misses by duplication

\subsection{SYCL Feature Evaluation}

\begin{figure}
         \centering
         \includegraphics[width=0.8\textwidth]{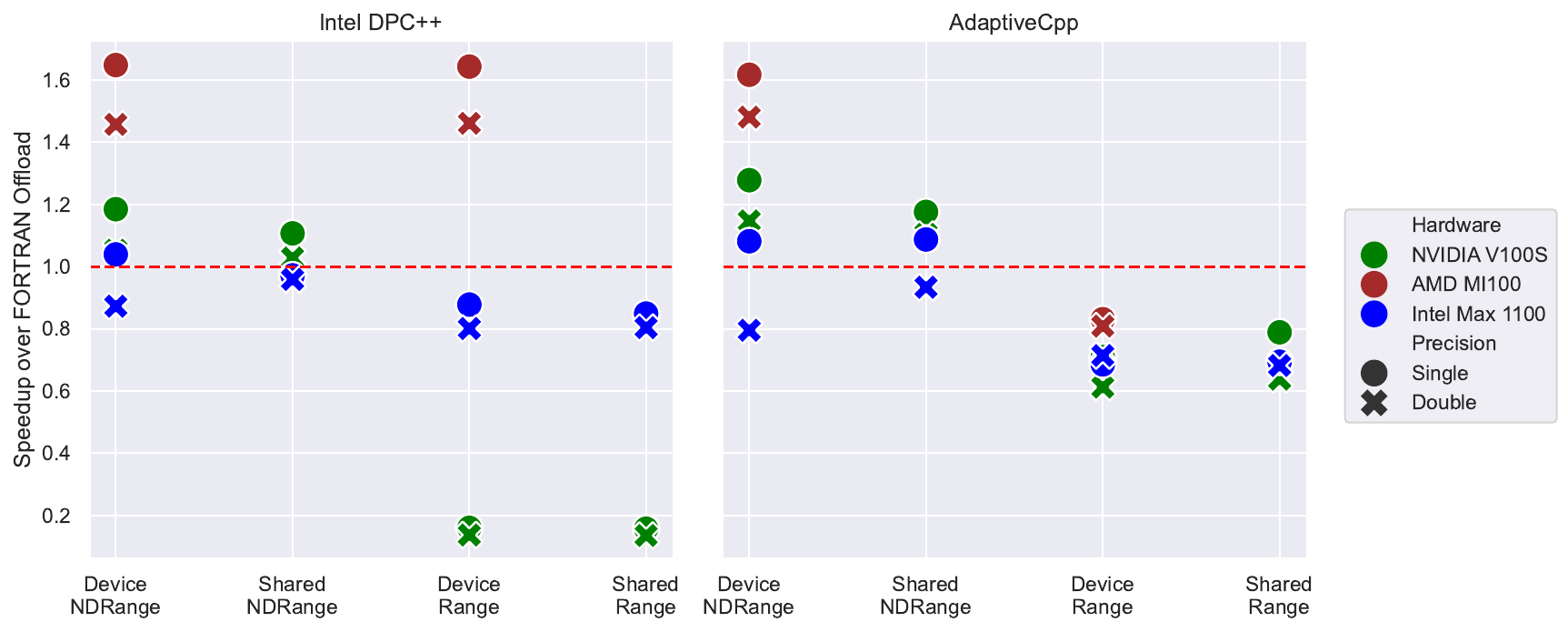}
         \label{fig:feature_eval_1}
        \caption{SYCL speedup over FORTRAN offload parallelism model by varying kernel and allocation type with column-major layout and in-order queues, on Lid-Driven Cavity}
        \label{fig:feature_eval}
\end{figure}

SYCL provides a great variety of built-in constructs to parallelize an application on heterogeneous hardware. However, the performance of each construct heavily depends on the target use case \cite{sycl-bench}.
% (e.g. shared allocation with numerous host-device transfers can severely reduce the bandwidth).
Furthermore, different SYCL platforms could implement the same feature in different way, adding additional complexity.
Figure \ref{fig:feature_eval} shows the performance of the Lid-Driven Cavity use case with each combination of USM allocations and SYCL kernel's type on every hardware and precisions, using both AdaptiveCpp and Intel DPC++. For \texttt{NDrange} kernels, the work group size is the biggest one possible on the target GPU (i.e. 1024 threads), organized as a block of 1x1024 threads. 
The values are normalized with the FORTRAN OpenMP offload backend, as it is the only one available on every hardware. In those benchmarks, we add a checkpoint at T = 50.000 to force data movement between host and device.

From the picture is clear that kernel performance in SYCL is heavily dependent on the adopted SYCL parallelization type. 
In particular, SYCL \texttt{range} are above the baseline in 33\% of the cases. On the other hand, \texttt{NDrange} kernels beat the baseline in 85\% of the configuration. However, some discrepancy between SYCL implementations arises: AdaptiveCpp \texttt{range} uses a work group size of 128 threads, organized in a 16x16 grid with 2-dimensional kernels. On NVIDIA GPUs,  the gridsize is divided into 256x256 blocks. This results in 15\% more uncoalasced global memory acccess compared to the \texttt{NDrange} version, where the work group size is unrolled along the y-axis (1x1024, or 1024x1 if in row-major).
Similar considerations apply also to other architectures.
On the other hand, DPC++ \texttt{range} kernels always select the largest possible work group size on GPU and put all the threads in one dimension (e.g. 1x1024 if the kernel is 2-dimensional). 
This heuristic performs well on both AMD and Intel, where \texttt{range} picks the same size as the one manually specified for \texttt{NDrange}. However, on NVIDIA GPUs, \texttt{range} kernels performance are detrimental, achieving only 15\% of the \texttt{NDrange} performance. The difference in performance between hardware is due to a small, but significant change in the work group size definition heuristic on Nvidia hardware: while on AMD and Intel hardware, threads are placed on the first dimension (e.g. 1024x1x1), on NVIDIA hardware DPC++ place the threads on the second dimension (1x1024x1). This results in 93\% more uncoalasced access on the Tesla V100S compared to the other hardware. By switching to a \texttt{row-major} layout, NVIDIA performance improves but it crashes performance on AMD and Intel GPUs for the same reason.
This discrepancy between work group size definitions severely limits the \texttt{range} performance portability across hardware.  
\newline
\newline
Regarding data management, shared allocations are not shown on AMD hardware: on AMD GPUs, on-demand page migration between host and device memory relies on the XNACK feature, which is disabled by default. However, XNACK is known to be experimental and unstable. When we enabled it, we encountered random kernel failures and GPU hangs. Consequently, we disabled it for this analysis.
Without XNACK enabled, \textit{shared} allocations function like \textit{host} allocations, meaning the data is allocated on the host and transferred to the device at each memory access, generating up to 1000x slowdown compared to the other implementation. Therefore, for this evaluation, we consider the AMD \textit{shared} backend as not available. 
On average, \texttt{device} allocation beat the baseline in 60\% of the configuration, while \texttt{shared} allocation only on 43\% of the cases. 
However, \texttt{shared} allocations performance depends on hardware support: for example, on NVIDIA GPU they beat the baseline 50\% of the time, while on the Intel GPU they only achieve better performance than the baseline on 37\%.
%It is worth noting that on Intel hardware, \texttt{shared} allocation are always prefetched on the device \cite{sycl-bench}, therefore  
%TODO finish
This variability, together with the unreliability of UVM on AMD, raises questions on its application to production scenarios.

% Come le performance sono in genere migliori su ndrange e range (Magari % di entry sotto e sopra la baseline)
% AdaptiveCpp in media peggio di DPC++, ma DPC++ ha gli estremi peggiori
% Meccanismo per scegliere kernel size in DPC++ e AdaptiveCPP
% Availability di Shared sugli HW (richiede UVM)
% Shared allocation performano similarly (poiche' l'applicazione carica i dati sulla GPU e poi li computa, non ci sono movimenti di dati out per quest exp (checkpointing disabilitato)
%slowdown medio di DPC++ range kernels vs AdaptiveCpp range kernels

\subsection{Performance Portability evaluation}

\textbf{\begin{table*}[!htb]
    \ra{1.3}
    \begin{minipage}{.49\linewidth}
      \centering
      \resizebox{0.95\columnwidth}{!}{%
      \begin{tabular}{@{}cllllrrrr@{}} \toprule
SYCL Impl.                & Hardware                      & USM & Kernel      & $AI$ & \makecell{$FR$ \\ \textit{(GFlop/s)}} & \makecell{$P$\\ \textit{(TFlop/s)}} & \makecell{\second{$e^{\prime}$}} \\ \midrule
\multirow{12}{*}{\rotatebox[origin=c]{90}{Intel DPC++}} & \multirow{5}{*}{NVIDIA V100S} & Shared               & Range & 0,38  & 125  & 0,43 &  29\% \\
                       &                               & Shared &  NDrange   & 1.37  & 976 & 1.55 &  63\% \\
                       &                               & Device & Range  & 0.38    & 124  & 0.43 & 29\%\\
                       &                               & Device & NDrange     & 1.27   & 934 & 1.55  & 60\%\\ \cdashline{2-8}
                       & \multirow{4}{*}{AMD MI100 (Est.)}   & Shared & Range    & X & X & X & X\\
                       &                               & Shared & NDrange & X  & X  & X   & X  \\
                       &                                & Device & Range    & 1.41 & 1132 & 1.60 & 70\%\\
                       &                               & Device & NDrange & 1.41  & 1132  & 1.60   & 70\%  \\ \cdashline{2-8}
                       & \multirow{4}{*}{Intel Max 1100}   & Shared & Range    & 1.22 & 781 & 0.976 & 80\%\\
                       &                               & Shared & NDrange & 1.22  & 782  & 0.976   & 80\%  \\
                       &                               & Device & Range  & 1.21   & 772 & 0.968   & 79\%  \\ 
                       &                               & Device & NDrange  & 1.21   & 775  & 0.968 & 80\% \\
\bottomrule
\end{tabular}%
}
\caption{\emph{miniLB} performance data for \textit{col\_MC} kernel,\\ single precision w/ Intel DPC++}
\label{tab:roofline-dpcpp}
\end{minipage}\hfill
    \begin{minipage}{.49\linewidth}
      \centering
      \resizebox{0.95\columnwidth}{!}{%
      \begin{tabular}{@{}cllllrrrr@{}} \toprule
SYCL Impl.                & Hardware                      & USM & Kernel      & $AI$ & \makecell{$FR$ \\ \textit{(GFlop/s)}} & \makecell{$P$\\ \textit{(TFlop/s)}} & \makecell{\second{$e^{\prime}$}} \\ \midrule
\multirow{12}{*}{\rotatebox[origin=c]{90}{AdaptiveCpp}} & \multirow{5}{*}{NVIDIA V100S} & Shared               & Range & 1,07  & 657  & 1.21 &  54\% \\
                       &                               & Shared &  NDrange   &  1.32 & 926 &  1.49 &  62\% \\
                       &                               & Device & Range  & 1.05   & 588  & 1.18 & 49\%\\
                       &                               & Device & NDrange     & 1.32 & 885  &  1.49 & 59.2\%\\ \cdashline{2-8}
                       & \multirow{4}{*}{AMD MI100 (Est.)}   & Shared & Range    & X & X & X & X\\
                       &                               & Shared & NDrange & X  & X  & X   & X  \\
                       &                               & Device & Range  & 1.14   & 601 & 1.38   & 44\%  \\ 
                       &                               & Device & NDrange          & 1.13   & 976  & 1.36 & 72\% \\ \cdashline{2-8}
                       & \multirow{4}{*}{Intel Max 1100}   & Shared & Range    & X & X & X & X\\
                       &                               & Shared & NDrange & 1.21  & 769  & 0.968   & 79\%  \\
                       &                               & Device & Range  &  1.04  & 603 & 0.83   & 72\%  \\ 
                       &                               & Device & NDrange &  1.21  & 900  & 0.968 & 93\% \\
\bottomrule
\end{tabular}%
}
\caption{miniLB performance data for \textit{col\_MC} kernel,\\ single precision w/ AdaptiveCpp}
\label{tab:roofline-acpp}
\end{minipage}
\end{table*}}

To measure the application performance portability, we employ the Pennycook \ppm metric \cite{pennycook2016metric}. However, we don't have a native optimized application version for each target hardware. In addition, calculating the application \textit{architectural efficiency} can be challenging, as it requires the identification of relevant bottlenecks on each hardware. For those reasons, to calculate the performance portability we employ the \textit{roofline efficiency}, which measures the distance between the application FLOP/s to the top of the roofline. Roofline efficiency has been demonstrated to successfully approximate architectural efficiency \cite{roofline_efficiency2}.
To calculate roofline efficiency, one needs to calculate the device peak performance 
\[ P  = \min \left( FR_m, BW_m \times AI_k \right) \]
where \( FR_m \) is the device floating point peak, $BW_m$ is the device bandwidth peak, and $AI_k$ is the arithmetic intensity for the application $k$, measured as the ratio between the application FLOP $FL$ and memory transferred. To capture those values, we used Nsight Compute on NVIDIA platform \cite{nsight} and Intel Advisor on Intel \cite{intel_advisor}. However, we encountered two difficulties.
%\begin{itemize}
%    \item 
    
First, the AMD MI100 does not provide FLOP counters, therefore it is not possible to calculate the application FLOP/s. However, \emph{miniLB} kernels do not have any device-dependent branch, therefore we expect the number of floating point operations to be the same on all three platform.
    To estimate the application FLOP/s, we use a similar methodology to the one defined in \cite{roofline_efficiency2}: \\
    % \begin{equation*}
    %     FR_{amd} = FR_{nvidia} * \frac{kernel\_time_{amd}}{kernel\_time_{nvidia}}
    % \end{equation*}
    $FR_{amd} = FR_{nvidia} * \frac{kernel\_time_{amd}}{kernel\_time_{nvidia}}$, 
    where $FR_{nvidia}$ is the floating point ratio of the corresponding implementation on NVIDIA hardware, and $\frac{kernel\_time_{amd}}{kernel\_time_{nvidia}}$ is the ratio between the two application kernel performance.

%    \item 
The second problem is related to AdaptiveCpp on Intel. Profiling an AdaptiveCpp-compiled application with Intel Advisor results in a profiler's internal exception, therefore we couldn't measure the flop rate on Intel Max. Depending on the kernel type, we followed different procedures:
    \begin{itemize}
        \item \texttt{NDrange}: Both DPC++ and AdaptiveCpp use the same work group size, therefore we approximate the flop rate by multiplying DPC++ flop rate by the ratio between the AdaptiveCpp and DPC++ performance.
        \item \texttt{range} Because AdaptiveCpp uses a different work group size compared to DPC++, we can't approximate the kernel bandwidth with high precision, therefore we skip this configuration. 
    \end{itemize}
\begin{figure}
    \centering
    \begin{minipage}{0.33\textwidth}
        \begin{subfigure}{\linewidth}
            \includegraphics[width=\textwidth]{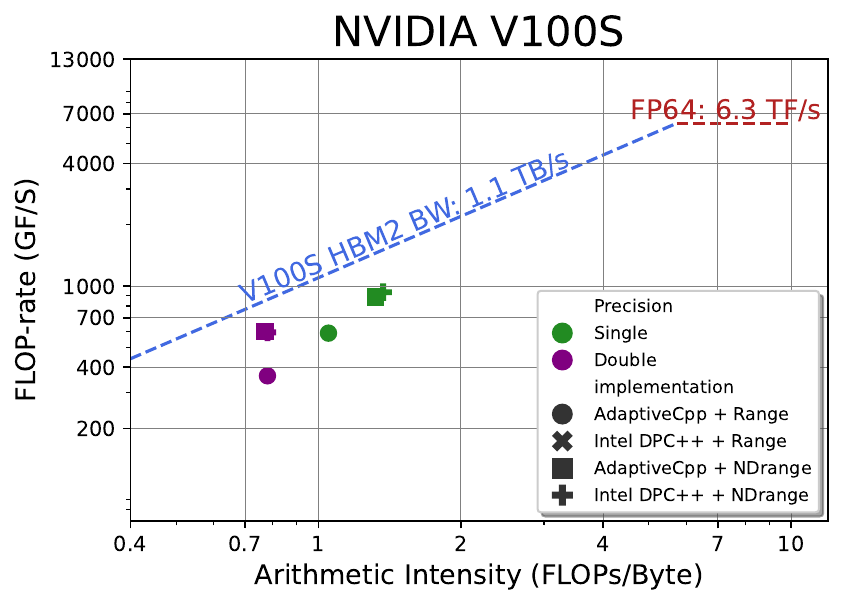}
        \end{subfigure}
        \begin{center}
        (A1) Device allocation
        \end{center}
        \begin{subfigure}{\linewidth}
            \includegraphics[width=\textwidth]{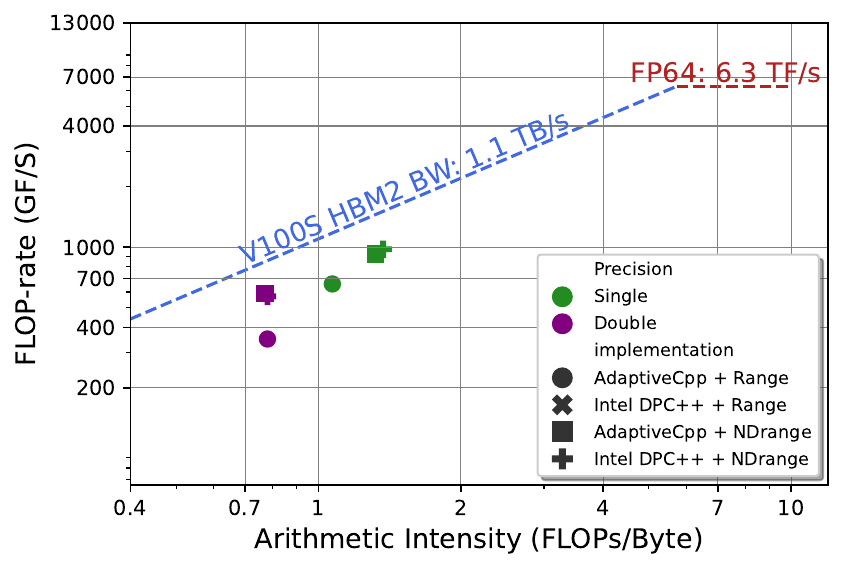}
        \end{subfigure}
        \begin{center}
        (A2) Shared allocation
        \end{center}
    \end{minipage} \hfill
    \begin{minipage}{0.33\textwidth}
        \begin{subfigure}{\linewidth}
            \includegraphics[width=\textwidth]{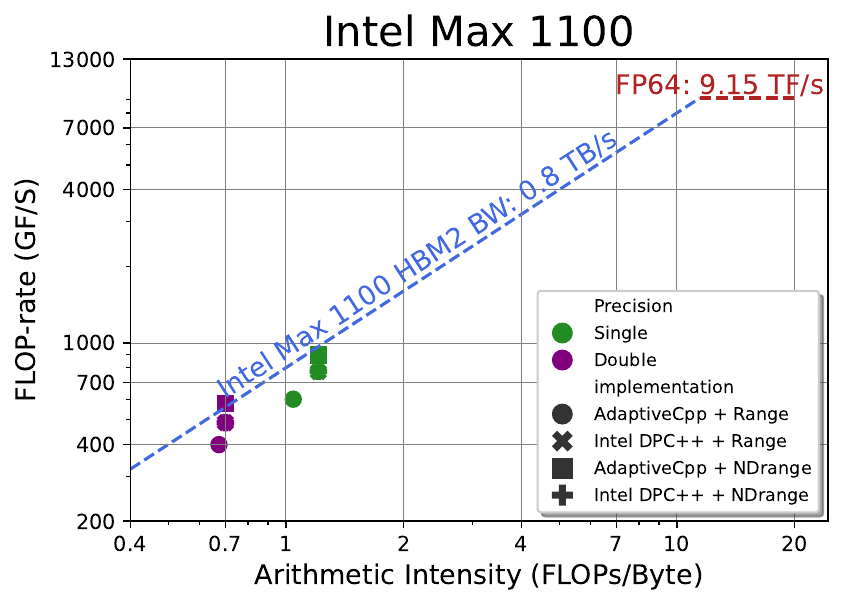}
        \end{subfigure}
        \begin{center}
        (B1) Device allocation
        \end{center}
        \begin{subfigure}{\linewidth}
            \includegraphics[width=\textwidth]{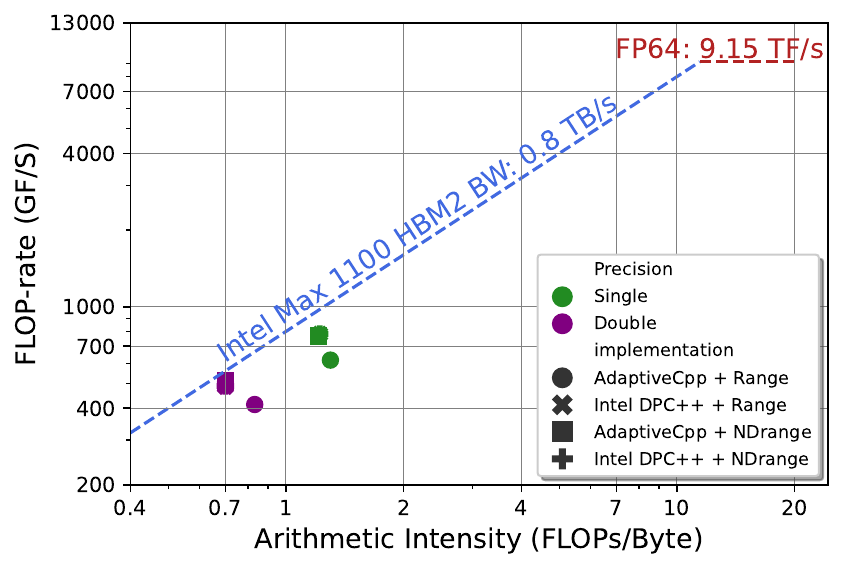}
        \end{subfigure}
        \begin{center}
        (B2) Device allocation
        \end{center}
    \end{minipage} \hfill
    \begin{minipage}{0.33\textwidth}
        \centering
        \begin{subfigure}{\linewidth}
            \includegraphics[width=\textwidth]{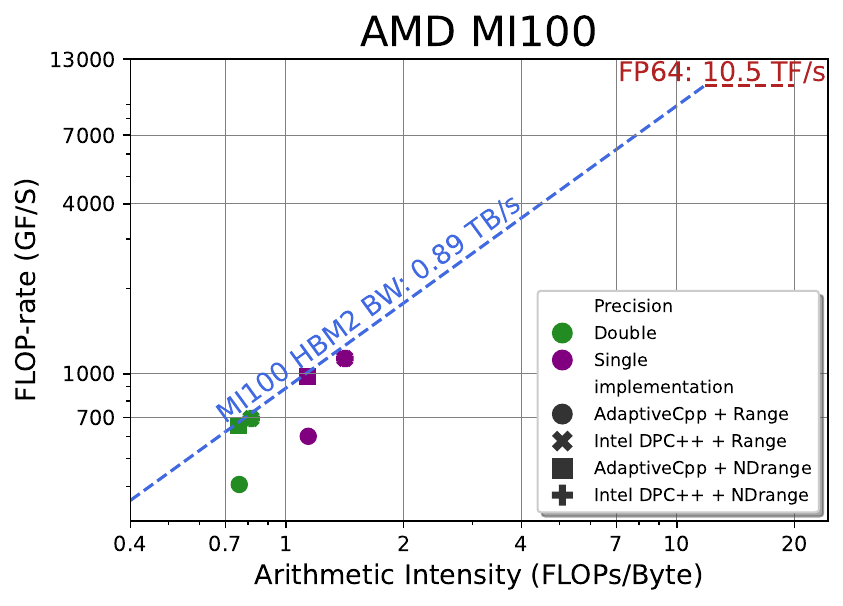}
        \end{subfigure}
        \begin{center}
        (C1) Device allocation
        \end{center}
    \end{minipage} 
    \caption{\emph{miniLB} roofline models per target hardware}
    \label{fig:roofline}
\end{figure}

\vspace{-1pt}
% For a fair analysis, we measured the maximum memory bandwidth of each GPU using the Empirical Roofline Toolkit (ERT) \cite{empirical-roofline-model}.
We measured the memory bandwidth of each hardware: our results showed a 1.1TB/s bandwidth on the NVIDIA V100S, 0.89TB/s on the AMD MI100, and 0.8TB/s on the Intel Max 1100.
Table \ref{tab:roofline-dpcpp}, \ref{tab:roofline-acpp}, shows the performance value collected for the fused collide and stream kernel, called \textit{col\_MC} kernel. $e^{\prime}$ indicates the distance from the roofline peak. The roofline results for both precision are shown in figure \ref{fig:roofline}.
As with every LB application, \emph{miniLB} is bandwidth-bound, therefore the device peak depends on the device bandwidth and arithmetic intensity.
For space constrain, we only show the results for single precision. Interestingly, \emph{miniLB} achieves at least 62\% of the device peak on every target hardware.
We can see that on Intel Max we achieve the highest roofline efficiency, getting up to 91\% of the peak on AdaptiveCpp with \texttt{NDrange}. 
While DPC++ and AdaptiveCpp show similar roofline efficiency for \texttt{NDrange} kernel, AdaptiveCpp \texttt{range} is 37\% and  8\% slower than DPC++ respectively on AMD and Intel \texttt{device} allocation. 
However, because of the previously mentioned uncoalesced access issue, DPC++ is 46\% and 42\% slower than AdaptiveCpp with \texttt{shared} and \texttt{device} allocation respectively on the NVIDIA V100S. This means that, while DPC++ could achieve better performance, on average AdaptiveCpp \texttt{range} heuristic is more portable among devices.
\newline

Finally, table \ref{tab:pp-1} and \ref{tab:pp-2} show the performance portability (\ppm) metric results for each precision. $\pp^{\prime}$ represents the value of performance portability considering only the current combination of data management backend and kernel type,
% while \ppm is the performance portability value considering the best implementation for each platform.
while \ppm is the maximum $\pp^{\prime}$ across all data management backend .
Because we treated \texttt{shared} allocation as not available on AMD hardware, \ppm is 0 for each \texttt{shared} configuration. 
\emph{miniLB} achieves a minimum of 60\% of performance portability among every precision, showing how SYCL can efficiently target any of the major vendor GPUs.
\texttt{NDrange} achieve a medium portability of 78\% among all precision, while \texttt{range} gets a medium portability of 62\%. It is worth noting that, while \texttt{NDrange} required a tuning phase to find the best work group size for each hardware, \texttt{range} achieved such results without any user intervention.

% \begin{table*}[!htb]
%     \ra{1.3}
%       \centering
%       \resizebox{0.35\columnwidth}{!}{%
%       \begin{tabular}{@{}cllrr@{}} \toprule
% Precision             & Kernel type                      & USM allocation & $\pp^{\prime}$      & $\pp$ \\ \midrule
% \multirow{4}{*}{\rotatebox[origin=c]{90}{Single}} & \multirow{2}{*}{Range} & Device & 64\% & \multirow{2}{*}{64\%} \\
% & & Shared & 0 &  \\ \cdashline{2-5}
% & \multirow{2}{*}{NDrange} & Device & 78\% & \multirow{2}{*}{78\%} \\ 
% & & Shared & 0 & \\ \cdashline{1-5}
% \multirow{4}{*}{\rotatebox[origin=c]{90}{Double}} & \multirow{2}{*}{Range} & Device &61\% & \multirow{2}{*}{61\%} \\
% & & Shared & 0 &  \\ \cdashline{2-5}
% & \multirow{2}{*}{NDrange} & Device & 78\% & \multirow{2}{*}{78\%} \\ 
% & & Shared & 0 & \\ 
% \bottomrule
% \end{tabular}%
% }
% \caption{\emph{miniLB} performance portability metric \\ for \textit{col\_MC} kernel for single and double precision}
% \label{tab:pp-1}
% \end{table*}

\begin{table*}[!htb]
\begin{minipage}{0.49\textwidth}
    
    \ra{1.3}
      \centering
      \resizebox{0.8\linewidth}{!}{%
      \begin{tabular}{@{}cllrr@{}} \toprule
Precision             & Kernel type                      & USM allocation & $\pp^{\prime}$      & $\pp$ \\ \midrule
\multirow{4}{*}{\rotatebox[origin=c]{90}{Single}} & \multirow{2}{*}{Range} & Device & 64\% & \multirow{2}{*}{64\%} \\
& & Shared & 0 &  \\ \cdashline{2-5}
& \multirow{2}{*}{NDrange} & Device & 78\% & \multirow{2}{*}{78\%} \\ 
& & Shared & 0 & \\
\bottomrule
\label{tab:pp-1}
\end{tabular}%
}
\caption{\emph{miniLB} performance portability metric \\ for \textit{col\_MC} kernel, single precision}
\label{tab:pp-1}
\end{minipage}
\hfill
\begin{minipage}{0.49\textwidth}
    
    \ra{1.3}
      \centering
      \resizebox{0.8\linewidth}{!}{%
      \begin{tabular}{@{}cllrr@{}} \toprule
Precision             & Kernel type                      & USM allocation & $\pp^{\prime}$      & $\pp$ \\ \midrule
\multirow{4}{*}{\rotatebox[origin=c]{90}{Double}} & \multirow{2}{*}{Range} & Device &61\% & \multirow{2}{*}{61\%} \\
& & Shared & 0 &  \\ \cdashline{2-5}
& \multirow{2}{*}{NDrange} & Device & 78\% & \multirow{2}{*}{78\%} \\ 
& & Shared & 0 & \\ 
\bottomrule
\end{tabular}%
}
\caption{\emph{miniLB} performance portability metric \\ for \textit{col\_MC} kernel, double precision}
\label{tab:pp-2}
\end{minipage}

\end{table*}

\vspace{-20pt}
\section{Conclusion and Future Work}
We presented \emph{miniLB}, the first, highly tunable, SYCL-based lattice Boltzmann mini-app. We successfully ported the original FORTRAN application to C++ and SYCL, achieving a considerable speedup on every platform. We analyzed a subset of the 96 possible \emph{miniLB} configuration settings to evaluate multiple combinations of SYCL features. 
We found that AdaptiveCpp and DPC++ portability can severely depend on the target SYCL feature, e.g. DPC++ \texttt{range} heuristic being less performance portable than AdaptiveCpp. Finally, we analyze \emph{miniLB} performance portability using the well-known \ppm metric. Our results show that \emph{miniLB} achieves high performance portability, with a \ppm value up to 78\%. \\
As a future work, we plan to implement more SYCL features and optimizations, e.g. local memory, \texttt{Buffer-Accessors}, specialization constants, etc., as well as extending the app to multi-GPU systems, using both low-level MPI calls and high-level SYCL frameworks like Celerity.
Furthermore, we would like to extend \emph{miniLB} to the 3D case and measure the impact of mixed precision computation in SYCL, for both numerical stability and energy consumption.

\begin{acknowledgments}
  This project has received funding from the Italian Ministry of University and Research under PRIN 2022 grant No. 2022CC57PY (LibreRT project).  
\end{acknowledgments}

\newpage

\bibliography{main}

\end{document}